\input{epsf}

\newcommand{\BOX}{\hbox {$\sqcap$ \kern -1em $\sqcup$}}

\newcounter{letter}

\documentclass {article} 
\usepackage {amsfonts}

\begin{document}

\title{The Meaning of Einstein's Equation}
\author{John C.\ Baez\footnote{
      Department of Mathematics, University of California,
      Riverside, California 92521, USA.
      email: baez@math.ucr.edu}
      \ and Emory F.\ Bunn\footnote{
      Physics Department, University of Richmond, Richmond, VA 23173, USA.
      email: ebunn@richmond.edu}}
\date{January 4, 2006}
\maketitle

\begin{abstract}
\noindent
This is a brief introduction to general relativity, designed for both
students and teachers of the subject.  While there are many excellent
expositions of general relativity, few adequately explain the
geometrical meaning of the basic equation of the theory: Einstein's
equation.  Here we give a simple formulation of this equation in terms
of the motion of freely falling test particles.  We also sketch some
of the consequences of this formulation and explain how it is
equivalent to the usual one in terms of tensors.  Finally, we include
an annotated bibliography of books, articles and websites suitable for
the student of relativity.

\end{abstract}

\section{Introduction}

General relativity explains gravity as the curvature of
spacetime.  It's all about geometry.  The basic equation
of general relativity is called Einstein's equation.   In units
where $c = 8 \pi G = 1$, it says
\begin{equation}
         G_{\alpha \beta} = T_{\alpha \beta} . 
\label{usualeinstein}\end{equation}
It looks simple, but what does it mean?   Unfortunately, the beautiful
geometrical meaning of this equation is a bit hard to find in most
treatments of relativity.  There are many nice popularizations that
explain the philosophy behind relativity and the idea of curved
spacetime, but most of them don't get around to explaining Einstein's
equation and showing how to work out its consequences.   There are
also more technical introductions which explain Einstein's equation in
detail --- but here the geometry is often hidden under piles of tensor 
calculus.

This is a pity, because in fact there is an easy way to express the
whole content of Einstein's equation in plain English.  In fact, after
a suitable prelude, one can summarize it in a single sentence!  One
needs a lot of mathematics to derive all the consequences of this
sentence, but it is still worth seeing --- and we can work out {\it
some} of its consequences quite easily.

In what follows, we start by outlining some differences between
special and general relativity.  Next we give a verbal formulation of
Einstein's equation.  Then we derive a few of its consequences
concerning tidal forces, gravitational waves, gravitational collapse,
and the big bang cosmology.  In the last section we explain why our
verbal formulation is equivalent to the usual one in terms of
tensors.  This article is mainly aimed at those who teach relativity,
but except for the last section, we have tried to make it accessible
to students, as a sketch of how the subject might be introduced.  We
conclude with a bibliography of sources to help teach the subject.

\section{Preliminaries}
\label{sec:prelim}

Before stating Einstein's equation, we need a little preparation.  We
assume the reader is somewhat familiar with special relativity ---
otherwise general relativity will be too hard.  But there are some big
differences between special and general relativity, which can cause
immense confusion if neglected.  

In special relativity, we cannot talk about {\it absolute} velocities,
but only {\it relative} velocities.  For example, we cannot sensibly ask
if a particle is at rest, only whether it is at rest relative to
another.  The reason is that in this theory, velocities are described as
vectors in 4-dimensional spacetime.  Switching to a different inertial
coordinate system can change which way these vectors point
relative to our coordinate axes, but not
whether two of them point the same way.

In general relativity, we cannot even talk about {\it relative}
velocities, except for two particles at the same point of spacetime ---
that is, at the same place at the same instant.  The reason is that in
general relativity, we take very seriously the notion that a vector is a
little arrow sitting at a particular point in spacetime.  To compare
vectors at different points of spacetime, we must carry one over to the
other.  The process of carrying a vector along a path without turning or
stretching it is called `parallel transport'.  When spacetime is
curved, the result of parallel transport from one point to another
depends on the path taken!  In fact, this is the very definition of what
it means for spacetime to be curved.  Thus it is ambiguous to ask whether
two particles have the same velocity vector unless they are at the same
point of spacetime.

It is hard to imagine the curvature of 4-dimensional spacetime, but
it is easy to see it in a 2-dimensional surface, like a sphere.  The
sphere fits nicely in 3-dimensional flat Euclidean space, so
we can visualize vectors on the sphere as `tangent vectors'.
If we parallel transport a tangent vector from the north pole to the
equator by going straight down a meridian, we get a different result
than if we go down another meridian and then along the equator:

\medskip
\centerline{\epsfysize=2.0in\epsfbox{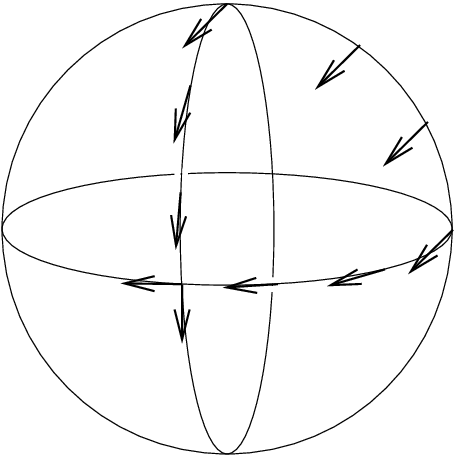}} \medskip

\noindent
Because of this analogy, in general relativity vectors are usually called
`tangent vectors'.  However, it is important not to take this 
analogy too seriously.  Our curved spacetime need not be embedded in some
higher-dimensional flat spacetime for us to understand its curvature, or
the concept of tangent vector.  The mathematics of tensor calculus is
designed to let us handle these concepts `intrinsically' --- i.e.,
working solely within the 4-dimensional spacetime in which we find
ourselves.   This is one reason tensor calculus is so important
in general relativity.

Now, in special relativity we can think of an inertial coordinate
system, or `inertial frame', as being defined by a field of 
clocks, all at rest relative to each other.   In general relativity
this makes no sense, since we can only unambiguously define the 
relative velocity of two clocks if they are at the same location.  
Thus the concept of inertial frame, so important in special relativity,
is {\it banned} from general relativity!   

If we are willing to put up with limited accuracy, we can still talk
about the relative velocity of two particles in the limit where they are
very close, since curvature effects will then be very small.   In
this approximate sense, we can talk about a  `local' inertial coordinate
system.  However, we must remember that this notion makes
perfect sense only in the limit where the region of spacetime covered by the
coordinate system goes to zero in size.  

Einstein's equation can be expressed as a statement about the relative
acceleration of very close test particles in free fall.  Let us clarify
these terms a bit.  A `test particle' is an idealized point particle
with energy and momentum so small that its effects on spacetime curvature
are negligible.  A particle is said to be in `free fall' when its motion
is affected by no forces except gravity.  In general relativity, a test
particle in free fall will trace out a `geodesic'.  This means that its
velocity vector is parallel transported along the curve it traces out
in spacetime.  A geodesic is the closest thing there is to a straight line 
in curved spacetime.  

Again, all this is easier to visualize in 2d space rather than 4d
spacetime.  A person walking on a sphere `following their nose' will
trace out a geodesic --- that is, a great circle.  Suppose two people
stand side-by-side on the equator and start walking north, both
following geodesics.  Though they start out walking parallel to each
other, the distance between them will gradually start to shrink, until
finally they bump into each other at the north pole.  If they didn't
understand the curved geometry of the sphere, they might think a `force'
was pulling them together.  

Similarly, in general relativity gravity is not really a `force', but
just a manifestation of the curvature of spacetime.  Note: not the
curvature of space, but of {\it spacetime}.  The distinction is
crucial.  If you toss a ball, it follows a parabolic path.  This is far from
being a geodesic in {\it space}: space is curved by the Earth's
gravitational field, but it is certainly not so curved as all that!  The
point is that while the ball moves a short distance in space, it moves
an enormous distance in {\it time}, since one second equals about
300,000 kilometers in units where $c = 1$.  This allows a slight amount
of spacetime curvature to have a noticeable effect.

\section{Einstein's Equation}

To state Einstein's equation in simple English, we need to consider a
round ball of test particles that are all initially at
rest relative to each other.  As we have seen, this is a sensible
notion only in the limit where the ball is very small. If we start with
such a ball of particles, it will, to second order in
time, become an ellipsoid as time passes.  This should not be too
surprising, because any linear transformation applied to a ball gives an
ellipsoid, and as the saying goes, ``everything is linear to first
order''.  Here we get a bit more: the relative velocity of the particles
starts out being zero, so to first order in time the ball does not
change shape at all: the change is a second-order effect.

Let $V(t)$ be the volume of the ball after a proper time $t$ has
elapsed, as measured by the particle at the center of the ball.  Then
Einstein's equation says: 
\[ {\ddot V\over V} \Bigr|_{t = 0} =
 -{1\over 2} \left( \begin{array}{l} 
{\rm flow \; of \;} t{\rm -momentum \; in \;\,} t {\rm \,\; direction \;} + \\ 
{\rm flow \; of \;} x{\rm -momentum \; in \;} x {\rm \; direction \;} + \\ 
{\rm flow \; of \;} y{\rm -momentum \; in \;} y {\rm \; direction \;} + \\ 
{\rm flow \; of \;} z{\rm -momentum \; in \;} z {\rm \; direction} 
\end{array} \right) \] 
where these flows are measured at the center of the ball at time zero,
using local inertial coordinates.  These flows are the diagonal
components of a $4 \times 4$ matrix $T$ called the `stress-energy
tensor'.  The components $T_{\alpha \beta}$ of this matrix say how much
momentum in the $\alpha$ direction is flowing in the $\beta$ direction
through a given point of spacetime, where $\alpha,\beta = t,x,y,z$.  The
flow of $t$-momentum in the $t$-direction is just the energy density,
often denoted $\rho$.  The flow of $x$-momentum in the $x$-direction is
the `pressure in the $x$ direction' denoted $P_x$, and similarly for $y$
and $z$.  It takes a while to figure out why pressure is really the flow
of momentum, but it is eminently worth doing.  Most texts explain this
fact by considering the example of an ideal gas.

In any event, we may summarize
Einstein's equation as follows:
\begin{equation}   {\ddot V\over V} \Bigr|_{t = 0} 
= -{1\over 2} (\rho + P_x + P_y + P_z). \label{einstein} \end{equation}
This equation says that positive energy density and positive pressure
curve spacetime in a way that makes a freely falling ball of point
particles tend to shrink.  Since $E = mc^2$ and we are working in units
where $c = 1$, ordinary mass density counts as a form of energy density.
Thus a massive object will make a swarm of freely falling particles at
rest around it start to shrink.  In short: {\it gravity attracts}.

We promised to state Einstein's equation in plain English, but have
not done so yet.  Here it is:

\begin{quote}
Given a small ball of freely falling test particles initially at rest
with respect to each other, the rate at which it begins to shrink is
proportional to its volume times: the energy density at the center of
the ball, plus the pressure in the $x$ direction at that point, plus
the pressure in the $y$ direction, plus the pressure in the $z$
direction.
\end{quote}

One way to prove this is by using the Raychaudhuri equation, discussions
of which can be found in the textbooks by Wald and by Ciufolini
and Wheeler cited in the bibliography.  But an elementary
proof can also be given starting from first principles,
as we will show in the final section of this article.

The reader who already knows some
general relativity may be somewhat skeptical that all 
of Einstein's equation is encapsulated in this formulation.  After
all, Einstein's equation in its usual tensorial form is really
a bunch of equations: the left and right sides of equation
(\ref{usualeinstein}) are $4\times 4$ matrices.  It is hard to believe
that the single equation (\ref{einstein}) captures all that
information.  It does, though, as long as we include one bit of fine
print: in order to get the full content of the Einstein equation from
equation (\ref{einstein}), we must consider small balls with 
{\it all possible} initial velocities --- i.e., balls that begin
at rest in all possible local inertial reference frames.

Before we begin, it is worth noting an even simpler formulation
of Einstein's equation that applies when the pressure is the
same in every direction:

\begin{quote}
Given a small ball of freely falling test particles initially at rest
with respect to each other, the rate at which it begins to shrink is
proportional to its volume times: the energy density at the center of
the ball plus three times the pressure at that point.
\end{quote}

\noindent
This version is only sufficient for `isotropic' situations: that
is, those in which all directions look the same in some local inertial
reference frame.  But, since the simplest models of cosmology treat
the universe as isotropic --- at least approximately, on large enough
distance scales --- this is all we shall need to derive an equation
describing the big bang!

\section{Some Consequences}

The formulation of Einstein's equation we have given is certainly not
the best for most applications of general relativity.   For example, in
1915 Einstein used general relativity to correctly compute the 
anomalous precession of the orbit of Mercury and also the deflection of
starlight by the Sun's gravitational field.  Both these calculations
would be very hard starting from equation (\ref{einstein}); they 
really call for the full apparatus of tensor calculus.   

However, we can easily use our formulation of Einstein's equation to get a 
qualitative --- and sometimes even quantitative  --- understanding of 
{\it some} consequences of general relativity.  We have already seen that it
explains how gravity attracts.  We sketch a few other consequences below.
These include Newton's inverse-square force law,
which holds in the limit of weak gravitational fields and small velocities,
and also the equations governing the big bang cosmology.

\subsubsection*{Tidal Forces, Gravitational Waves}

We begin with some qualitative consequences of Einstein's equation.
Let $V(t)$ be the volume of a small ball of test particles in free fall
that are initially at rest relative to each other.  
In the vacuum there is no energy density or
pressure, so $\ddot V|_{t = 0} = 0$, but the curvature of spacetime can
still distort the ball.  For example, suppose you drop a small ball of
instant coffee when making coffee in the morning.   The grains of 
coffee closer to the earth accelerate towards it a bit more, causing the ball 
to start stretching in the vertical direction.  However, as the grains all
accelerate towards the center of the earth, the ball also starts
being squashed in the two horizontal directions.  Einstein's equation says
that if we treat the coffee grains as test particles, these two effects
cancel each other when we calculate the second derivative of the ball's
volume, leaving us with $\ddot V|_{t = 0} = 0$.    It is a fun exercise
to check this using Newton's theory of gravity!   

This stretching/squashing of a ball of falling coffee grains is an
example of what people call `tidal forces'.  As the name suggests, another
example is the tendency for the ocean to be stretched in one direction
and squashed in the other two by the gravitational pull of the moon.  

Gravitational waves are another example of how spacetime can be curved
even in the vacuum.   General relativity predicts that when any heavy 
object wiggles, it sends out ripples of spacetime curvature which 
propagate at the speed of light.  This is far from obvious starting from
our formulation of Einstein's equation!  It also predicts
that as one of these ripples of curvature passes by, our small ball of
initially test particles will be stretched in one transverse
direction while being squashed in the other transverse direction.   
From what we have already said, these effects must precisely cancel 
when we compute $\ddot V|_{t = 0}$.  

Hulse and Taylor won the Nobel prize in 1993 for careful observations of
a binary neutron star which is slowly spiraling down, just as general
relativity predicts it should, as it loses energy by emitting 
gravitational radiation.  Gravitational waves have not been {\it
directly} observed, but there are a number of projects underway to
detect them.  For example, the LIGO project will bounce a laser between
hanging mirrors in an L-shaped detector, to see how one leg of the
detector is stretched while the other is squashed.  Both legs are 
4 kilometers long, and the detector is designed to be sensitive
to a $10^{-18}$-meter change in length of the arms.  

\subsubsection*{Gravitational Collapse}

Another remarkable feature of Einstein's equation is the pressure term: it 
says that not only energy density but also pressure causes gravitational
attraction.  This may seem to violate our intuition that pressure makes
matter want to expand!  Here, however, we are talking about {\it
gravitational} effects of pressure, which are undetectably small in
everyday circumstances. To see this, let's restore the factors
of $c$.  Also, let's remember that in ordinary circumstances
most of the energy is in the form of rest energy, so we can
write the energy density $\rho$ as $\rho_m c^2$, 
where $\rho_m$ is the ordinary mass
density:
\[   {\ddot V\over V}\Bigr|_{t = 0} 
= -{4\pi G}(\rho_m + {1\over c^2}(P_x + P_y + P_z)). \]
On the human scale all of the terms on the right are small, 
since $G$ is very small.  (Gravity is a weak force!)
Furthermore, the pressure terms are much smaller than the mass density
term, since they are divided by an extra factor of $c^2$. 

There are a number of important situations in which $\rho$ does not
dominate over $P$.  In a neutron star, for example, which is held up
by the degeneracy pressure of the neutronium it consists of, pressure
and energy density contribute comparably to the right-hand side of
Einstein's equation.  Moreover, above a mass of about 2 solar masses a
nonrotating neutron star will inevitably collapse to form a black
hole, thanks in part to the gravitational attraction caused by
pressure.  In fact, any object of mass $M$ will form a black hole if
it is compressed to a radius smaller than its Schwarzschild radius,
$R = 2GM/c^2$.

\subsubsection*{Newton's Inverse-Square Force Law}

A basic test of general relativity is to check that it reduces to 
good old Newtonian gravity in the limit where gravitational effects are 
weak and velocities are small compared to the speed of light.
To do this, we can use our formulation of Einstein's equation 
to derive Newton's inverse-square force law for a planet with mass $M$ 
and radius $R$.   Since we can only do this when gravitational effects 
are weak, we must assume that the planet's radius is much greater than 
its Schwarzschild radius: $R \gg M$ in units where $c = 8\pi G = 1$.  
Then the curvature of {\it space} -- as opposed to spacetime -- 
is small.   To keep things simple, we make a couple of additional 
assumptions: the planet has uniform density $\rho$, and the pressure 
is negligible. 
 
We want to derive the familiar Newtonian result 
\[ 
a = -{GM\over r^2} 
\]
giving the radial gravitational acceleration $a$ of a test particle 
at distance $r$ from the planet's center, with $r>R$ of course.  
 
To do this, let $S$ be a sphere of radius $r$ centered on the planet.  Fill 
the interior of the sphere with test particles, all of which are initially 
at rest relative to the planet.  At first, this might seem like an 
illegal thing to do: we know that notions like `at rest' only make sense 
in an infinitesimal neighborhood, and $r$ is not infinitesimal.  But 
because space is nearly flat for weak gravitational fields, we can 
get away with this. 
 
We can't apply our formulation of Einstein's equation directly to  
$S$, but we can apply it to any infinitesimal sphere within 
$S$.  In the picture below, the solid black circle represents the 
planet, and the dashed circle is $S$.  The interior of $S$ has been 
divided up into many tiny spheres filled with test particles.  Green 
spheres are initially inside the planet, and red spheres are outside. 
 
\centerline{\epsfxsize 5in \epsfbox{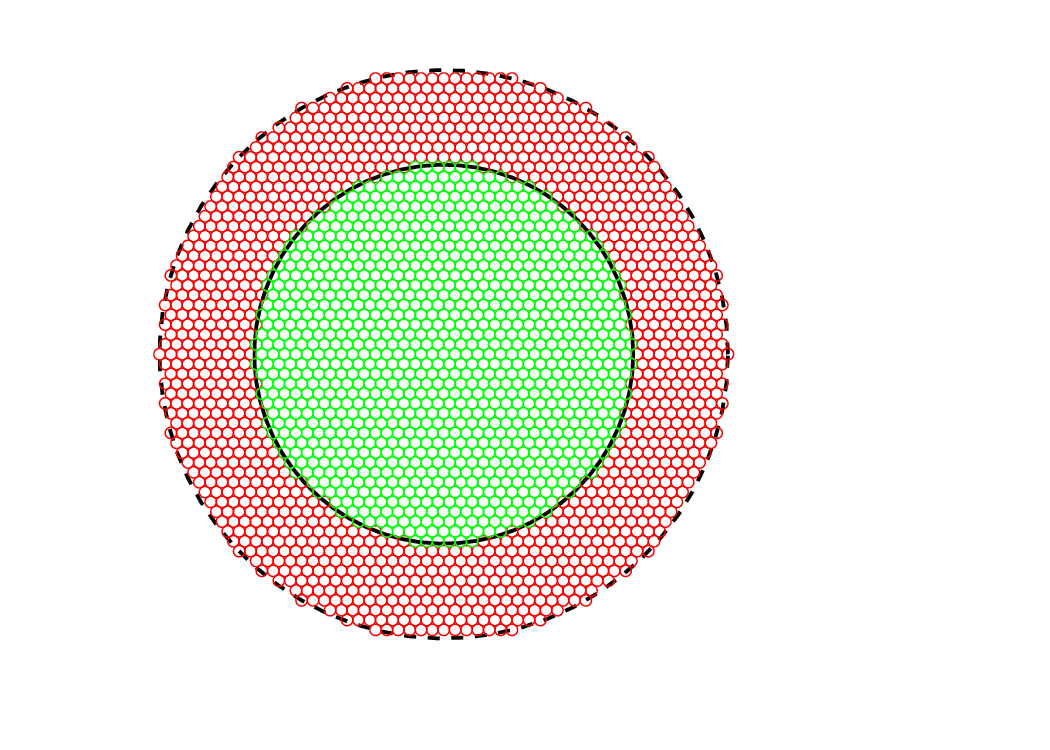}} 

Suppose we pick a tiny green sphere that lies within the planet's 
volume, a distance less than $R$ from the center.   Our formulation 
of Einstein's equation tells us that the fractional change in volume 
of this sphere will be  
\[ 
\left.{\ddot V\over V}\right|_{t=0}=-{1\over 2}\rho 
\qquad\qquad\mbox{(inside the planet).} 
\]
On the other hand, as we saw in our discussion of tidal effects,  
spheres outside the planet's volume will be distorted in shape by 
tidal effects, but remain unchanged in volume.  So, any little red 
sphere that lies outside the planet will undergo no change in volume 
at all: 
\[ 
\left.{\ddot V\over V}\right|_{t=0}=0\qquad\qquad\mbox{(outside the  
planet).} 
\]
Thus, after a short time $\delta t$ has elapsed, the test 
particles will be distributed like this: 
 
\centerline{\epsfxsize 5in \epsfbox{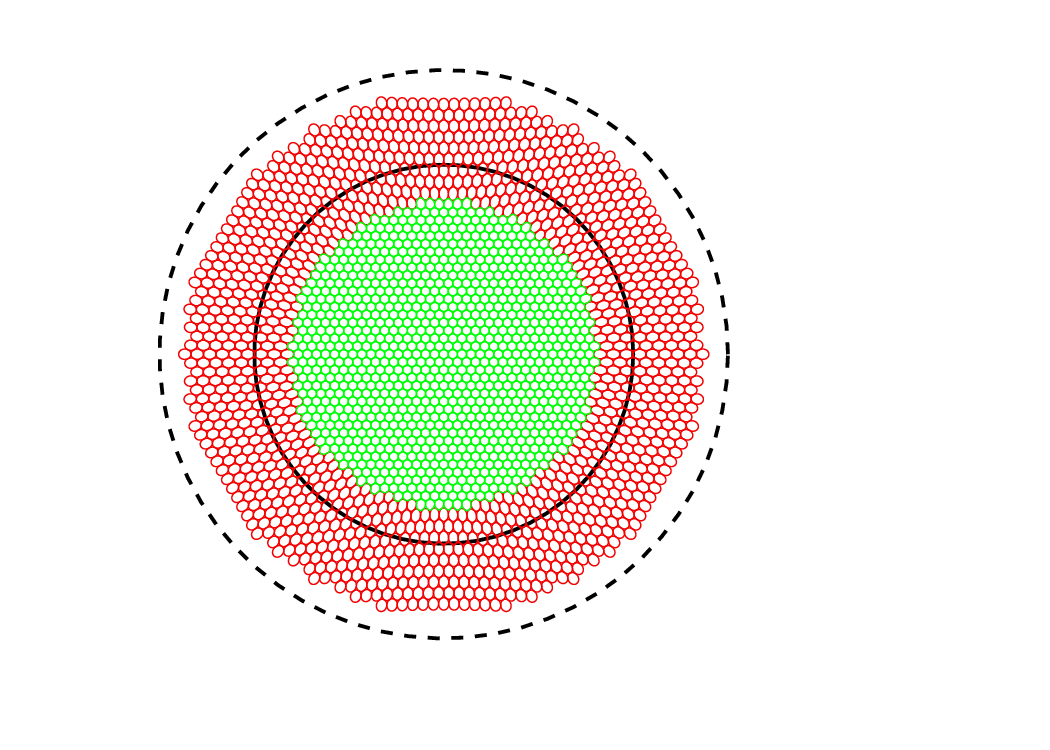}} 
 
Because the volume occupied by the spheres of particles within 
the planet went down, the whole big sphere of test particles had 
to shrink.  Now consider the following question:  
what is the change in volume of the large sphere of test 
particles?  All the little green spheres shrank by the same fractional 
amount:
\[ 
{\delta V\over V}
= {1\over 2}\left(\ddot V\over V\right)(\delta t)^2
= - {1\over 4}\rho (\delta t)^2, 
\] 
to leading order in $\delta t$, 
while all the little red spheres underwent no change in volume. 
So, if $V_P$ stands for the volume occupied by the test particles 
that were initially inside the planet, and $V_{S}$ stands for the
volume of the big sphere $S$, then the overall change in volume is
\begin{eqnarray*} 
\delta V_S &=& \delta V_P             \\ 
&=&\left(\delta V\over V\right)V_P    \\ 
&=& -{1\over 4}\rho(\delta t)^2 \, V_P \\ 
&=&-{1\over 4}M(\delta t)^2. 
\end{eqnarray*} 

We can also express $\delta V_S$ in terms of the change in 
the sphere's radius, $\delta r$: 
\[ 
\delta V_S = 4\pi r^2 \delta r. 
\]
Setting the two expressions for $\delta V_S$ equal, we see
\[
\delta r = -{M\over 16\pi r^2}(\delta t)^2. 
\]
But the change in radius $\delta r$ is just the radial displacement of 
a test particle on the outer edge of the sphere, so
$\delta r={1\over 2} a (\delta t)^2$ where $a$ is the radial
acceleration of this particle.
Plugging this into the above equation, we obtain
\[
a = -{M\over 8\pi r^2}. 
\]
We've been working all along in units in which $8\pi G=1$.  Switching 
back to conventional units, we obtain
\[ 
a = -{GM\over r^2}
\] 
just as Newton said.
 
\subsubsection*{The Big Bang}

We can also derive
some basic facts about the big bang cosmology.  Let us assume the
universe is not only expanding but also homogeneous and isotropic.
The expansion of the universe is vouched for by the redshifts of
distant galaxies.  The other assumptions also seem to be approximately
correct, at least when we average over small-scale inhomogeneities 
such as stars and galaxies.  For simplicity, we will imagine the
universe is homogeneous and isotropic even on small scales.

An observer at any point in such a universe would see all objects
receding from her.  Suppose that, at some time $t=0$, she identifies a
small ball $B$ of test particles centered on her.  Suppose this ball
expands with the universe, remaining spherical as time passes because
the universe is isotropic.  Let $R(t)$ stand for the radius of this ball 
as a function of time.  The Einstein equation will give us an equation of
motion for $R(t)$.  In other words, it will say how the
expansion rate of the universe changes with time.

It is tempting to apply equation (\ref{einstein}) to the ball $B$, but
we must take care.  This equation applies to a ball of
particles that are initially at rest relative to one another --- that
is, one whose radius is not changing at $t=0$.  However, the ball $B$
is expanding at $t=0$.  Thus, to apply our formulation of Einstein's
equation, we must introduce a second small ball of test particles
that are at rest relative to each other at $t = 0$.  

Let us call this second ball $B'$, and call its radius as a function
of time $r(t)$.  Since the particles in this ball begin at rest
relative to one another, we have
\[    \dot r(0) = 0.  \]
To keep things simple, let us also assume that at $t=0$ both balls
have the exact same size:
\[     r(0) = R(0)  .\]

Equation (\ref{einstein}) applies to the ball $B'$, since
the particles in {\it this} ball are initially at rest relative to
each other.  Since the volume of this ball is proportional to $r^3$, 
and since $\dot r = 0$ at $t=0$, the left-hand side of 
equation (\ref{einstein}) is simply
\[
\left.{\ddot V\over V}\right|_{t=0}=\left.{3\ddot r\over r}\right|_{t=0}.
\]
Since we are assuming the universe is isotropic, we know that the 
various components of pressure are equal: 
$P_x = P_y = P_z = P$.  Einstein's equation (\ref{einstein}) thus says 
that 
\[ \left.{3\ddot r\over r}\right|_{t=0} = -{1\over 2}(\rho + 3 P)  .\]

We would much prefer to rewrite this expression in terms of $R$ 
rather than $r$.  Fortunately, we can do this.  
At $t=0$, the two spheres have the same radius: $r(0)=R(0)$.  Furthermore, 
the second derivatives are the same: $\ddot r(0) = \ddot R(0)$.  This 
follows from the equivalence principle, which says that, at any given 
location, particles in free fall do not accelerate with respect to each 
other.  At the moment $t=0$, each test particle on the surface of
the ball $B$ is right next to a corresponding test particle in $B'$.  
Since they are not accelerating with respect to each other, the observer 
at the origin must see both particles accelerating in the same way,
so $\ddot r(0) = \ddot R(0)$.
It follows that we can replace $r$ with $R$ in the above equation,
obtaining
\[ \left.{3\ddot R\over R}\right|_{t=0} = -{1\over 2}(\rho + 3 P)  .\]

We derived this equation for a very small ball, but
in fact it applies to a ball of any size.  This is because, in a
homogeneous expanding universe, the balls of all radii
must be expanding at the same fractional rate.  In other words,
$\ddot R/R$ is independent of the radius $R$, although
it can depend on time.  Also, there is nothing special in
this equation about the moment $t=0$, so the equation must
apply at all times.  In summary, therefore, the basic equation
describing the big bang cosmology is
\begin{equation}  {3\ddot R\over R} = -{1\over 2}(\rho + 3 P),
\label{bang}
\end{equation}
where the density $\rho$ and pressure $P$ can depend on time but
not on position.  Here we can imagine $R$ to be the separation 
between any two `galaxies'. 

To go further, we must make more assumptions about the nature of
the matter filling the universe.  One simple model is a universe
filled with pressureless matter.  Until recently, this
was thought to be an accurate model of our universe.  Setting $P=0$, we 
obtain 
\[  {3\ddot R\over R} = -{\rho \over 2} . \]
If the energy density of the universe is mainly due to the mass
in galaxies, `conservation of galaxies' implies that $\rho R^3 = k$ 
for some constant $k$.  This gives
\[  {3\ddot R\over R} = - {k \over 2 R^3}  \]
or
\[   \ddot R = -{k \over 6 R^2}  .\]
Amusingly, this is the same as the equation of motion for a particle in
an attractive $1/R^2$ force field.  In other words, the equation 
governing this simplified cosmology is the same as the Newtonian
equation for what happens when you throw a ball vertically upwards from
the earth!  This is a nice example of the unity of physics.  Since
``whatever goes up must come down --- unless it exceeds escape velocity,''
the solutions of this equation look roughly like this:

\medskip
\centerline{\epsfysize=2.0in\epsfbox{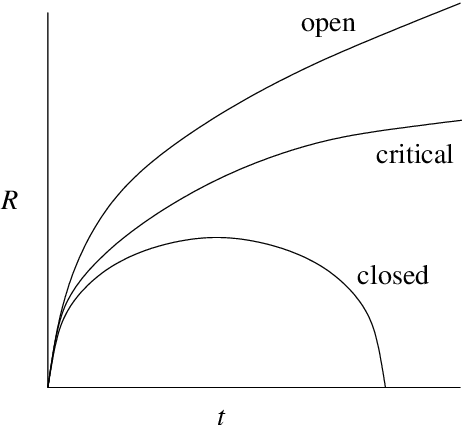}} \medskip

\noindent
So, the universe started out with a big bang!   It will
expand forever if its current rate of expansion is sufficiently
high compared to its current density, but it will recollapse in a 
`big crunch' otherwise.   

\subsubsection*{The Cosmological Constant}

The simplified big bang model just described is inaccurate for
the very early history of the universe, when the pressure of radiation
was important.  Moreover, recent observations seem to indicate that it is
seriously inaccurate even in the present epoch.  First of all, it seems
that much of the energy density is not accounted for by known forms of
matter.  Still more shocking, it seems that the expansion of the
universe may be accelerating rather than slowing down!  One possibility
is that the energy density and pressure are nonzero even for the vacuum.
For the vacuum to not pick out a preferred notion of `rest', its 
stress-energy tensor must be proportional to the metric.  In local
inertial coordinates this means that the stress-energy tensor
of the vacuum must be
\[ T = \left( \begin{array}{cccc} \Lambda & 0 &0& 0 \\ 
                                 0 & -\Lambda &0 & 0 \\
                                 0 & 0 &-\Lambda & 0 \\
                                 0 & 0 &0 & -\Lambda \\ 
\end{array}\right) \] 
where $\Lambda$ is called the `cosmological
constant'.  This amounts to giving empty space an energy density equal
to $\Lambda$ and pressure equal to $-\Lambda$, so that $\rho + 3P$ for
the vacuum is $-2\Lambda$.  Here pressure effects dominate because there are
more dimensions of space than of time!  If we add this cosmological
constant term to equation (\ref{bang}), we get 
\begin{equation}
{3\ddot R\over R} = -{1\over 2}(\rho + 3 P - 2\Lambda) ,
\label{lambdaequation}
\end{equation}
where $\rho$ and $P$ are the energy density and pressure due to matter.
If we treat matter as we did before, this gives
\[ {3\ddot R\over R} = -{k \over 2R^3} + \Lambda . \] 
Thus, once the universe expands sufficiently, the cosmological constant
becomes more important than the energy density of matter in
determining the fate of the universe.  If $\Lambda > 0$, 
a roughly exponential expansion will then ensue.  This seems
to be happening in our universe now.

\subsubsection*{Spatial Curvature}

We have emphasized that gravity is due not just to the 
curvature of space, but of {\it spacetime}.  
In our verbal formulation of Einstein's equation, this shows
up in the fact that we consider particles moving forwards in time 
and study how their paths deviate in the space directions.  
However, Einstein's equation also gives information about
the curvature of space.  To illustrate this, it is easiest
to consider not an expanding universe but a static one.

When Einstein first tried to use general relativity to construct a
model of the entire universe, he assumed that the universe must be
static --- although he is said to have later described this as ``his
greatest blunder''.  As we did in the previous section, Einstein
considered a universe containing ordinary matter with density $\rho$,
no pressure, and a cosmological constant $\Lambda$.  Such a universe
can be static --- the galaxies can remain at rest with respect to each
other --- only if the right-hand side of equation
(\ref{lambdaequation}) is zero.  In such a universe, the cosmological
constant and the density must be carefully `tuned' so that
$\rho=2\Lambda$.  It is tempting to conclude that spacetime in this
model is just the good old flat Minkowski spacetime of special
relativity.  In other words, one might guess that there are no
gravitational effects at all.  After all, the right-hand side of
Einstein's equation was tuned to be zero.  This would be a mistake,
however.  It is instructive to see why.

Remember that equation (\ref{einstein}) contains all the information
in Einstein's equation only if we consider all possible small balls.
In all of the cosmological applications so far, we have applied the
equation only to balls whose centers were at rest with respect to the
local matter.  It turns out that only for such balls is the
right-hand side of equation (\ref{einstein}) zero in the
Einstein static universe.

To see this, consider a small ball of test particles, initially at
rest relative to each other, that is moving with respect to the matter
in the universe.  In the local rest frame of such a ball, the
right-hand side of equation (\ref{einstein}) is nonzero.  For one
thing, the pressure due to the matter no longer vanishes.  Remember
that pressure is the flux of momentum.  In the frame of our moving
sphere, matter is flowing by.  Also, the energy density goes up, both
because the matter has kinetic energy in this frame and because of
Lorentz contraction.  The end result, as the reader can verify, is
that the right-hand side of equation (\ref{einstein}) is negative for
such a moving sphere.  In short, although a stationary ball of test
particles remains unchanged in the Einstein static universe, our
moving ball shrinks!

This has a nice geometric interpretation: the geometry in this model
has spatial curvature.  As we noted in section \ref{sec:prelim}, on a
positively curved surface such as a sphere, initially parallel lines
converge towards one another.  The same thing happens in the
three-dimensional space of the Einstein static universe.  In
fact, the geometry of space in this model is that of a 3-sphere.
This picture illustrates what happens:  

\medskip
\centerline{\epsfysize=2.0in\epsfbox{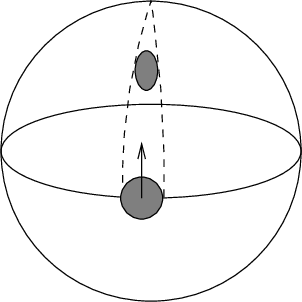}} \medskip

\noindent
One dimension is suppressed in this
picture, so the two-dimensional spherical surface shown represents the
three-dimensional universe.  The small shaded circle on the
surface represents our tiny sphere of test particles, which starts at
the equator and moves north.  The sides of the sphere approach each
other along the dashed geodesics, so the sphere shrinks in the
transverse direction, although its diameter in the direction of motion
does not change.

As an exercise, the reader who wants to test his understanding can fill in the
mathematical details in this picture and determine the
radius of the Einstein static universe in terms of the density.
Here are step-by-step instructions:

\begin{itemize}

\item Imagine an observer moving at speed $v$ through a cloud of
stationary particles of density $\rho$.  Use special relativity
to determine the energy density and pressure in the observer's
rest frame.  Assume for simplicity that the observer is
moving fairly slowly, so keep only the lowest-order nonvanishing
term in a power series in $v$.

\item Apply equation (\ref{einstein}) to a sphere in this 
frame, including the contribution due to the cosmological
constant (which is the same in all reference frames).
You should find that the volume of the sphere decreases
with $\ddot V/V\propto -\rho v^2$ to leading order in $v$.

\item Suppose that space in this universe has the geometry of a large
3-sphere of radius $R_U$.  Show that the radii in the directions
transverse to the motion start to shrink at a rate given by 
$\left.(\ddot R/R)\right|_{t=0}
= -v^2/R_U^2$.  (If, like most people, you are better at visualizing
2-spheres than 3-spheres, do this step by considering a small circle
moving on a 2-sphere, as shown above, rather than a small sphere
moving on a 3-sphere.  The result is the same.)

\item Since our little sphere is shrinking in two dimensions, its
volume changes at a rate $\ddot V/V = 2\ddot R/R$.  Use Einstein's
equation to relate the radius $R_U$ of the universe to the
density $\rho$.

\end{itemize}

The final answer is $R_U = \sqrt{2/\rho}$.  As you can confirm in standard
textbooks, this is correct.

Spatial curvature like this shows up in the expanding cosmological
models described earlier in this section as well.  In principle,
the curvature radius can be found from our formulation of Einstein's
equation by similar reasoning in these expanding models.  In fact,
however, such a calculation is extremely messy.  Here the apparatus of
tensor calculus comes to our rescue.  

\section{The Mathematical Details}

To see why equation (\ref{einstein}) is equivalent to the usual
formulation of Einstein's equation, we need a bit of tensor calculus.
In particular, we need to understand the Riemann curvature tensor
and the geodesic deviation equation.  For a detailed explanation
of these, the reader must turn to some of the texts in the bibliography.
Here we briefly sketch the main ideas.

When spacetime is curved, the result of parallel transport depends on
the path taken.  To quantify this notion, pick two vectors $u$ and $v$
at a point $p$ in spacetime.  In the limit where $\epsilon \to 0$, we
can approximately speak of a `parallelogram' with sides $\epsilon u$ and
$\epsilon v$.  Take another vector $w$ at $p$ and parallel transport it
first along $\epsilon v$ and then along $\epsilon u$ to the opposite
corner of this parallelogram.  The result is some vector $w_1$.
Alternatively, parallel transport $w$ first along $\epsilon u$ and then
along $\epsilon v$.  The result is a slightly different vector, $w_2$:

\medskip
\centerline{\epsfysize=1.8in\epsfbox{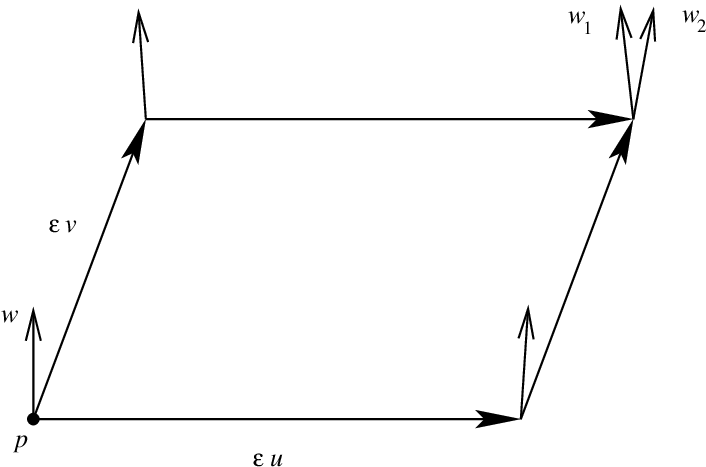}} \medskip

\noindent 
The limit 
\begin{equation}
   \lim_{\epsilon \to 0} {w_2 - w_1 \over \epsilon^2} = R(u,v)w 
\label{riemann} \end{equation}
is well-defined, and it measures the curvature of spacetime at the point
$p$. In local coordinates we can write it as
\[    R(u,v)w = R^\alpha_{\beta\gamma\delta} u^\beta v^\gamma w^\delta,\]
where as usual we sum over repeated indices.  The quantity 
$R^\alpha_{\beta\gamma\delta}$ is called the `Riemann curvature
tensor'.  

We can use this tensor to compute the relative acceleration
of nearby particles in free fall if they are initially at rest
relative to one another.  Consider
two freely falling particles at nearby points $p$ and $q$.  Let $v$ be
the velocity of the particle at $p$, and let $\epsilon u$ be the vector
from $p$ to $q$.  Since the two particles start out at rest relative
to one other, the velocity of the particle at $q$ is obtained by 
parallel transporting $v$ along $\epsilon u$.

Now let us wait a short while.  Both particles trace out geodesics as
time passes, and at time $\epsilon$ they will be at new points, say $p'$
and $q'$.  The point $p'$ is displaced from $p$ by an amount $\epsilon v$,
so we get a little parallelogram, exactly as in
the definition of the Riemann curvature:

\medskip
\medskip
\centerline{\epsfysize=1.4in\epsfbox{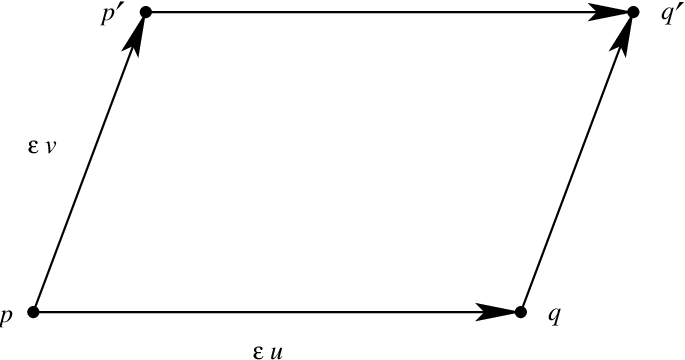}} \medskip

\noindent
Next let us compute the new relative velocity
of the two particles.  To compare vectors we must carry one to
another using parallel transport.  Let $v_1$ be the vector we get by
taking the velocity vector of the particle at $p'$ and parallel
transporting it to $q'$ along the top edge of our parallelogram.  Let
$v_2$ be the velocity of the particle at $q'$.  The difference $v_2 -
v_1$ is the new relative velocity.  Here is a picture of the whole
situation:

\medskip
\centerline{\epsfysize=1.8in\epsfbox{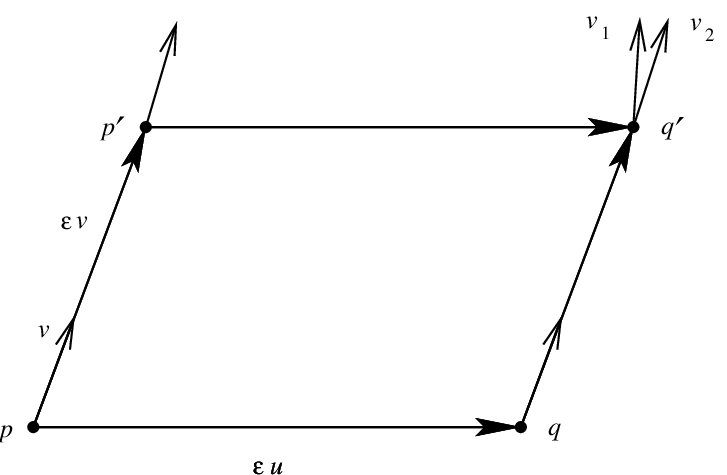}} \medskip

\noindent
The vector $v$ is depicted as shorter than $\epsilon v$ for
purely artistic reasons.  

It follows that over this passage of time, the average
relative acceleration of the two particles is 
$a = (v_2 - v_1)/\epsilon$.  By equation (\ref{riemann}), 
\[  \lim_{\epsilon \to 0} {v_2 - v_1 \over \epsilon^2} = R(u,v)v,\]
so 
\[  \lim_{\epsilon \to 0} {a \over \epsilon} = R(u,v)v . \]
This is called the `geodesic deviation equation'.  From the
definition of the Riemann curvature it is easy to see that
$R(u,v)w = -R(v,u)w$, so we can also write this equation as
\begin{equation} \lim_{\epsilon \to 0} {a^\alpha \over \epsilon} =
 - R^\alpha_{\beta\gamma\delta} v^\beta u^\gamma v^\delta .
\label{deviation}
\end{equation}

Using this equation we can work out the second time derivative of the
volume $V(t)$ of a small ball of test particles that start out at rest
relative to each other.  As we mentioned earlier, to second order
in time the ball changes to an ellipsoid.   Furthermore, since the ball 
starts out at rest, the principal axes of this ellipsoid don't rotate
initially.   We can therefore adopt local inertial coordinates in which, 
to second order in $t$, the center of the ball is at rest and the three
principal axes of the ellipsoid are aligned with the three spatial
coordinates.  Let $r^j(t)$ represent the radius of the $j$th axis 
of the ellipsoid as a function of time.  If the ball's initial radius 
is $\epsilon$, then
$$
r^j(t) = \epsilon + {1\over 2}a^j t^2 + O(t^3),
$$
or in other words,
$$
\lim_{t\to 0} {\ddot r^j\over r^j} = \lim_{t\to 0}{a^j\over\epsilon}.
$$
Here the acceleration $a^j$ is given by equation (\ref{deviation}), with
$u$ being a unit vector in the $j$th coordinate direction
and $v$ being the velocity of the ball, which is a unit vector
in the time direction.
In other words,
$$
\lim_{t\to 0} {\ddot r^j(t)\over r^j(t)} = - R^j_{\beta j\delta}v^\beta 
v^\delta=-R^j_{tjt}.
$$
No sum over $j$ is implied in the above expression.

Since the volume of our ball is proportional to the product of the radii,
$\ddot V/V \to \sum_j \ddot r^j/r^j$ as $t\to 0$, so
\[    \lim_{V \to 0} {\ddot V\over V}\Bigr|_{t = 0} = 
 - R^\alpha_{t\alpha t} , \]
where now a sum over $\alpha$ is implied.  The sum over $\alpha$
can range over all four coordinates, not just the
three spatial ones, since the symmetries of
the Riemann tensor demand that  $R^{t}_{ttt}=0$.

The right-hand side is minus the time-time component 
of the `Ricci tensor'
\[      R_{\beta\delta} =  R^\alpha_{\beta\alpha\delta}. \]
That is,
%\[    \lim_{V \to 0} {\ddot V\over V}  \Bigr|_{t = 0} = 
% - R_{\beta\delta} v^\beta v^\delta .\]
%In local inertial coordinates where the ball starts out
%at rest we have $v = (1,0,0,0)$, so 
\begin{equation} 
 \lim_{V \to 0} {\ddot V\over V}\Bigr|_{t = 0} = 
 - R_{tt} \label{contract} \end{equation}
in local inertial coordinates where the ball starts out at rest.

In short, the Ricci tensor says how our ball of freely 
falling test particles starts changing in volume.  The Ricci
tensor only captures some of the information in the Riemann curvature
tensor.  The rest is captured by something called the `Weyl tensor',
which says how any such ball starts changing in shape.  The Weyl tensor
describes tidal forces, gravitational waves and the like.

Now, Einstein's equation in its usual form says 
\[   G_{\alpha \beta} = T_{\alpha \beta} .\]
Here the right side is the stress-energy tensor, while the left side,
the `Einstein tensor', is just an abbreviation for a
quantity constructed from the Ricci tensor:
\[    G_{\alpha \beta} = R_{\alpha \beta} - {1\over 2}g_{\alpha \beta}
R^\gamma_\gamma. \]
Thus Einstein's equation really says
\begin{equation}   R_{\alpha \beta} - {1\over 2}g_{\alpha \beta}
R^\gamma_\gamma  = T_{\alpha \beta} . \label{einstein1} \end{equation}
This implies 
\[   R^\alpha_\alpha -  {1\over 2}g^\alpha_\alpha
R^\gamma_\gamma  = T^\alpha_\alpha, \]
but $g^\alpha_\alpha = 4$, so 
\[   -R^\alpha_\alpha = T^\alpha_\alpha .\]
Plugging this in equation (\ref{einstein1}), we get
\begin{equation}   R_{\alpha \beta} = T_{\alpha \beta} 
- {1\over 2}g_{\alpha \beta}  T^\gamma_\gamma 
\label{einstein2}. \end{equation}
This is equivalent version of Einstein's equation, 
but with the roles of $R$ and $T$ switched!  The good thing about
this version is that it gives a formula for the Ricci tensor, which
has a simple geometrical meaning.

Equation (\ref{einstein2}) will be true if any one component holds in
all local inertial coordinate systems.  This is a bit like the
observation that all of Maxwell's equations are contained in Gauss's law
and $\nabla \cdot B = 0$.  Of course, this is only true if we
know how the fields transform under change of coordinates.  Here we
assume that the transformation laws are known.  Given this, Einstein's
equation is equivalent to the fact that 
\begin{equation}   R_{tt} = T_{tt} - {1\over 2}g_{tt} T^\gamma_\gamma 
\label{einstein3} \end{equation}
in every local inertial coordinate system about every point.  In
such coordinates we have 
\begin{equation}  g =  \left( \begin{array}{cccc} 
                 -1 & 0 &0 & 0  \\ 
                 0     & 1 &0 & 0  \\
                 0     & 0 &1 & 0  \\
                 0     & 0 &0 & 1  \\
\end{array}\right)
\label{standard}
\end{equation}
so $g_{tt} = -1$ and 
\[   T^\gamma_\gamma = -T_{tt} + T_{xx} + T_{yy} + T_{zz}. \]
Equation (\ref{einstein3}) thus says that
\[   R_{tt} = {1\over 2} (T_{tt} + T_{xx} + T_{yy} + T_{zz}) .\]
By equation (\ref{contract}), this is equivalent to
\[  \lim_{V \to 0} {\ddot V\over V}\Bigr|_{t = 0} = 
- {1\over 2} (T_{tt} + T_{xx} + T_{yy} + T_{zz}). \]
As promised, this is the simple, tensor-calculus-free formulation 
of Einstein's
equation.

\section{Bibliography}

\subsection*{Websites}

There is a lot of material on general relativity available online,
including many books, such as this:

\medskip
\noindent
{\it Lecture Notes on General Relativity}, S.\ M.\ Carroll, \hfill
\break http://arxiv.org/abs/gr-qc/9712019

\medskip 
\noindent
The free online journal {\it Living Reviews in Relativity} is
an excellent way to learn more about many aspects of relativity.
One can access it at:

\medskip
\noindent
{\sl Living Reviews in Relativity}, http://www.livingreviews.org

\medskip
\noindent
Part of learning relativity is working one's way through certain classic
confusions.  The most common are dealt with in the ``Relativity
and Cosmology'' section of this site:

\medskip 
\noindent
{\it Frequently Asked Questions in Physics}, 
edited by D.\ Koks
\hfill \break http://math.ucr.edu/home/baez/physics/

\subsection*{Nontechnical Books}

Before diving into the details of general relativity, it is good
to get oriented by reading some less technical books.  Here are
four excellent ones written by leading experts on the subject:

\medskip
\noindent
{\it General Relativity from A to B}, R.\ Geroch
(University of Chicago Press, Chicago, 1981).

\medskip
\noindent
{\it Black Holes and Time Warps: Einstein's Outrageous Legacy}, K.\ S.\ 
Thorne (W.\ W.\ Norton, New York, 1995).

\medskip
\noindent
{\it Gravity from the Ground Up:
An Introductory Guide to Gravity and General Relativity},
B.\ F.\ Schutz (Cambridge University Press, Cambridge 2003).

\medskip
\noindent
{\it Space, Time, and Gravity: the Theory of the Big Bang and Black Holes,} 
R.\ M.\ Wald (University of Chicago Press, Chicago, 1992).

\subsection*{Special Relativity}

Before delving into general relativity in a more technical way, one
must get up to speed on special relativity.  Here are two excellent
texts for this:

\medskip
\noindent
{\it Introduction to Special Relativity}, W.\ Rindler
(Oxford University Press, Oxford, 1991).

\medskip
\noindent
{\it Spacetime Physics: Introduction to Special Relativity},
E.\ F.\ Taylor and J.\ A.\ Wheeler
(W.\ H.\ Freeman, New York, 1992).

\subsection*{Introductory Texts}

When one is ready to tackle the details of general relativity,
it is probably good to start with one of these textbooks:

\medskip
\noindent
{\it Introducing Einstein's Relativity}, 
R.\ A.\ D'Inverno
(Oxford University Press, Oxford, 1992).

\medskip
\noindent
{\it Gravity: An Introduction to Einstein's General Relativity},
J.\ B.\ Hartle (Addison-Wesley, New York, 2002).

\medskip
\noindent
{\it Introduction to General Relativity},
L.\ Hughston and K.\ P.\ Tod
(Cambridge University Press, Cambridge, 1991).

\medskip
\noindent
{\it A First Course in General Relativity},
B.\ F.\ Schutz
(Cambridge University Press, Cambridge, 1985).

\medskip
\noindent
{\it General Relativity: An Introduction to the 
Theory of the Gravitational Field},
H.\ Stephani
(Cambridge University Press, Cambridge, 1990).

\subsection*{More Comprehensive Texts}

To become a expert on general relativity, one really must tackle
these classic texts:

\medskip
\noindent 
{\it Gravitation}, 
C.\ W.\ Misner, K.\ S.\ Thorne, and J.\ A.\ Wheeler
(W.\ H.\ Freeman, New York, 1973).

\medskip
\noindent 
{\it General Relativity},
R.\ M.\ Wald
(University of Chicago Press, Chicago, 1984).

\medskip
\noindent
Along with these textbooks, you'll want to do lots of problems!
This book is a useful supplement:

\medskip
\noindent 
{\it Problem Book in Relativity and Gravitation}, A. Lightman and R.\ H.\ 
Price (Princeton University Press, Princeton, 1975).

\subsection*{Experimental Tests}

The experimental support for general relativity up to the early
1990s is summarized in:

\medskip
\noindent 
{\it Theory and Experiment in Gravitational Physics}, Revised Edition,
C.\ M.\ Will
(Cambridge University Press, Cambridge, 1993).

\medskip
\noindent
A more up-to-date treatment of the subject can be found in:

\medskip
\noindent 
``The Confrontation between General Relativity and Experiment,''
C.\ M.\ Will, 
{\it Living Reviews in Relativity} 4,  4 (2001). 
Available online at:
\hfill \break
http://www.livingreviews.org/lrr-2001-4

\subsection*{Differential Geometry}

The serious student of general relativity will experience a
constant need to learn more tensor calculus --- or in modern
terminology, `differential geometry'.  Some of this can be found
in the texts listed above, but it is also good to read
mathematics texts.  Here are a few:

\medskip
\noindent 
{\it Gauge Fields, Knots and Gravity},
J.\ C.\ Baez and J.\ P.\ Muniain
(World Scientific, Singapore, 1994).

\medskip
\noindent 
{\it An Introduction to Differentiable Manifolds and Riemannian Geometry}, 
W.\ M.\ Boothby
(Academic Press, New York, 1986).

\medskip
\noindent
{\it Semi-Riemannian Geometry with Applications to Relativity}, 
B.\ O'Neill
(Academic Press, New York, 1983).

\subsection*{Specific Topics}

The references above are about general relativity as a whole.  Here
are some suggested starting points for some of the particular topics
touched on in this article.

\paragraph{The meaning of Einstein's equation.}
Feynman gives a quite different approach to this in:

\medskip
\noindent
{\it The Feynman Lectures on Gravitation}, R.\ P.\ Feynman et al.\ 
(Westview Press, Boulder, Colorado, 2002).

\medskip
\noindent
His approach focuses on the curvature of space rather than the
curvature of spacetime.

\paragraph{The Raychaudhuri equation.}  This equation, which
is closely related to our formulation of Einstein's equation,
is treated in some standard textbooks, including the one by Wald mentioned
above.  A detailed discussion can be found in

\medskip\noindent
{\it Gravitation and Inertia}, I.\ Ciufolini and J.\ A.\ Wheeler (Princeton
University Press, Princeton, 1995).

\paragraph{Gravitational waves.}  

Here are two nontechnical descriptions of the binary pulsar work
for which Hulse and Taylor won the Nobel Prize:

\medskip\noindent ``The Binary Pulsar: Gravity Waves Exist,''
C.\ M.\ Will, {\it Mercury}, Nov-Dec 1987, p.\ 162.

\medskip\noindent
``Gravitational Waves from an Orbiting Pulsar,'' J.\ M.\ Weisberg, 
J.\ H.\ Taylor, and L.\ A.\ Fowler, 
{\it Scientific American}, Oct 1981, p.\ 74.

\medskip
\noindent
Here is a review article on the ongoing efforts to directly detect
gravitational waves:

\medskip
\noindent
``Detection of Gravitational Waves,'' J.\ Lu, D. G.\ Blair, and C.\ 
Zhao, {\it Reports on Progress in Physics}, 63, 1317-1427 (2000).

\medskip
\noindent
Some present and future experiments to detect gravitational radiation are
described here:

\medskip
\noindent
{\it LIGO Laboratory Home Page},
http://www.ligo.caltech.edu/

\medskip
\noindent
{\it The Virgo Project},
http://www.virgo.infn.it/

\medskip
\noindent
{\it Laser Interferometer Space Antenna}, http://lisa.jpl.nasa.gov/

\paragraph{Black holes.}
Astrophysical evidence that black holes exist is summarized in:

\medskip
\noindent
``Evidence for Black Holes,'' M.\ C.\ Begelman, {\it Science}, 300, 1898-1903
(2003).

\medskip
\noindent
A less technical discussion of the particular case of the supermassive
black hole at the center of our Milky Way Galaxy can be found here:

\medskip
\noindent
{\it The Black Hole at the Center of Our Galaxy}, F.\ Melia (Princeton
University Press, Princeton, 2003).

\paragraph{Cosmology.}  There are lots of good popular books on cosmology.
Since the subject is changing rapidly, pick one that is up to date.
At the time of this writing, we recommend:

\medskip
\noindent
{\it The Extravagant Universe: Exploding Stars, Dark Energy, and the 
Accelerating Cosmos}, R.\ P.\ Kirshner (Princeton University Press, 
Princeton, 2002).

\medskip
\noindent
A good online source of cosmological information is:

\medskip\noindent
{\it Ned Wright's Cosmology Tutorial}, 
\hfill \break http://www.astro.ucla.edu/$\sim$wright/cosmolog.htm

\medskip

\noindent
The following cosmology textbooks are arranged in increasing order
of technical difficulty:

\medskip
\noindent
{\it Cosmology: The Science of the Universe}, 2nd ed., E.\ Harrison
(Cambridge University Press, Cambridge, 2000).

\medskip
\noindent
{\it Cosmology: a First Course}, M.\ Lachi\`eze-Rey (Cambridge
University Press, Cambridge, 1995).

\medskip
\noindent
{\it Principles of Physical Cosmology}, P.\ J.\ E.\ Peebles (Princeton
University Press, Princeton, 1993).

\medskip
\noindent
{\it The Early Universe}, E.\ W.\ Kolb and M.\ S.\ Turner (Addison--Wesley,
New York, 1990).

\medskip
\noindent
{\it The Large-Scale Structure of Spacetime}, 
S.\ W.\ Hawking and G.\ F.\ R.\ Ellis (Cambridge University Press, 
Cambridge, 1975).

\section*{Acknowledgment}

E.F.B. is supported by National Science Foundation grant 0233969.

\end{document}